\documentclass[twocolumn,aps,10pt,prl]{revtex4-1}

\def\bwt{\begin{widetext}}

\def\ewt{\end{widetext}}

\def\be{\begin{equation}}

\def\ee{\end{equation}}

\def\bea{\begin{eqnarray}}

\def\eea{\end{eqnarray}}

\def\bean{\begin{eqnarray*}}

\def\eean{\end{eqnarray*}}

\def\bary{\begin{array}}

\def\eary{\end{array}}

\def\bit{\begin{itemize}}

\def\eit{\end{itemize}}

\usepackage{graphicx}
\usepackage{hyperref}
\usepackage{times}

\begin{document}

\title{The Generalized Abelian and Non-Abelian Gauge Theories and Their Particle Physics Applications}

\author{Tianjun Li}

\affiliation{
Key Laboratory of Theoretical Physics, Institute of Theoretical Physics,
Chinese Academy of Sciences, Beijing 100190, China
}
\affiliation{
School of Physical Sciences, University of Chinese Academy of Sciences,
No.~19A Yuquan Road, Beijing 100049, China
}



\begin{abstract}

Gauge theory is the foundation of the particle physics Standard Model (SM).
Considering the multiple gauge sectors for one gauge transformation, 
we study the generalized Abelian and non-Abelian (Yang-Mills theory) 
gauge theories. We first point out that the $U(1)$ gauge theory has 
a few unique properties, which provide the motivations for the generalized 
Yang-Mills theory. Also, we consider the generalized Abelian gauge theory, 
and study the Higgs mechanism with new interesting properties. In addition, 
we propose the simple and generic generalizations of Yang-Mills theory. 
In the simple generalization, we realize two specific properties in the Abelian 
gauge theory.  For applications in particle physics, we propose 
the invisible axion model with TeV-scale Peceei-Quinn symmetry breaking.
We can solve the strong CP problem, 
and obtain the effective decay constant around the intermediate scale. Moreover,
we study the SM electroweak symmetry breaking induced from the symmetry breaking 
in the other gauge Sector. In particular, we can easily obtain 
the strong first order electroweak phase transition in the SM, which is important 
for electroweak baryogenesis and gravitational wave.

\end{abstract}
\maketitle


{\bf Introduction~--}~There are four fundamental interactions in the Nature: stong interaction,
weak interaction, electromagentic interaction (or hypercharge interaction), and gravity. 
The first three fundamental interactions are described by the gauge theories with
groups $SU(3)_C$, $SU(2)_L$, and $U(1)_Y$ respectively in the 
Standard Model (SM). In particular, the SM particle content has been confirmed
since the SM-like Higgs boson $(h)$ with mass $m_h=125.09\pm 0.24$~GeV has been
 discovered at the LHC~\cite{ATLAS, CMS}.
However, the SM has some problems and then cannot be the final theory. 
Thus, we need to explore the new physics beyond the SM.
And there are many possible directions:
supersymmetry, extra gauge symmetries, dark matter, composite Higgs fields, 
neutrino masses and mixing from seesaw mechanism,  extra dimensions,
and Grand Unified Theory (GUT), etc. 

In this paper, we shall take a completely different approach, and  
study the generalizations of the Abelian and non-Abelian (or say the Yang-Mills theory) gauge theories,
which are the foundations of the SM gauge interactions.
The generalized Yang-Mills theories have been studied before as follows: 
(i) The role of gauge fields is not only played by vector fields,
but also scalar and tensor fields~\cite{Chaves:1998hg, Chaves:1999qu, Chaves:2000qe, Castro:2006kd}; 
(ii) If the base space is non-commutative, the
algebra of gauge transformations is a mixture of the finite-dimensional Lie algebra and the
algebra of functions on the non-commutative space~\cite{Ho:2015bia}. Here, we propose a brand new scenario:
there are multiple gauge Sectors for one gauge transformation.
First, we consider the Abelian $U(1)$ gauge theory, and compare the differences between 
the Abelian gauge theory and Yang-Mills theory. We point out that the $U(1)$ gauge theory has four
unique properties, which provide the motivations for 
the generalized Yang-Mills theory. Also, we consider the generalized Abelian gauge theory,
and study the Higgs mechanism for the model with two gauge Sectors.
If gauge symmetry is broken in one Sector, the gauge symmetry in the other
sector will be broken naturally via the tadpole term or trilinear term.
In particular, the induced symmetry breaking can be strong first order.
Second, we consider the generalized Yang-Mills theories. 
In the simple generalization of the non-Abelian gauge theory 
with muliple gauge Sectors for the same gauge transformation, 
we realize two specific properties in the Abelian gauge theory. In addition, we propose the
generic generalization of the Yang-Mills theory, which cannot realize any specific property
in the Abelian gauge theory. However, how to construct the non-trivial gauge invariant interactions 
between different gauge sectors is a big challenge. 
Third, we study the applications of the generalized gauge theories in particle physics.
We propose the viable invisible axion model~\cite{Peccei:1977hh, Weinberg:1977ma, Wilczek:1977pj,
Kim:1979if, Shifman:1979if, Dine:1981rt, Zhitnitsky:1980tq}, whose Peceei-Quinn (PQ) symmetry
is broken around the TeV scale. Our axion model
 can still solve the strong CP problem, and has the effective decay constant around 
intermediate scale. Moreover, we study the SM electroweak 
symmetry breaking induced from the symmetry breaking in the other gauge Sector. In particular,
we can easily obtain the strong first order electroweak phase transition in the SM, which is important for 
electroweak baryogenesis and gravitational wave. Furthemore, we point out that
the deconstruction of extra dimensions~\cite{ArkaniHamed:2001ca, Hill:2000mu},
 mirror models~\cite{Lee:1956qn, Kobzarev:1966qya, Pavsic:1974rq, Kolb:1985bf, Foot:1991bp},
top color models~\cite{Topcolor}, top flavour models~\cite{Topflavor}, 
and top hypercharge models~\cite{Chiang:2007sf}, etc, can be realized by the
generalized Abelian and non-Abelian gauge theories with multiple gauge sectors.
And our proposal might be applied to the heterotic $E_8\times E_8$
string theory as well~\cite{Gross:1984dd, Gross:1985fr, Horava:1995qa, Horava:1996ma}.

{\bf The Abelian Gauge Theory and Its generalizations~--}~We shall consider the $U(1)$ gauge theory
and introduce the scalar particles $\Phi_i$ and fermionic particles $\Psi_i$ with $U(1)$ charges
$q_{\Phi_i}$ and $q_{\Psi_i}$, respectively. The relevant Lagrangian is given by
\begin{eqnarray}
{\cal L} &=& -\frac{1}{4} F^{\mu\nu} F_{\mu\nu} 
+ i\overline{\Psi}_{i} \gamma^{\mu} D_{\mu} \Psi_{i}+|D_{\mu} \Phi_i|^2 ~,~
\end{eqnarray}
where the gauge field strength is
$F_{\mu\nu} = \partial_{\mu} A_{\nu} -\partial_{\nu} A_{\mu}$, and the covariant deriavative is 
$D_{\mu} = \partial_{\mu} -ig q_X A_{\mu}$ where $q_X$ is the $U(1)$ charge 
of the particle $X= \Phi_i/\Psi_i$.
Without loss of generality, one can choose one particle with $U(1)$ charge equal to 1
as a normalization. In this paper, we shall not take such normalization and 
consider the general formalism. 
The Lagrangian is invariant under the following gauge transformation
\begin{eqnarray}
&& \Phi_i \rightarrow U^{q_{\Phi_i}} \Phi_i~,~~\Psi_i \rightarrow U^{q_{\Psi_i}} \Psi_i~,~~\nonumber \\
&& A_{\mu} \rightarrow U A_{\mu} U^{-1} + \frac{i}{g} U\partial_{\mu}U^{-1}
= A_{\mu} + \frac{\partial_{\mu} \Lambda}{g}~,~
\end{eqnarray}
where $U=e^{i\Lambda(x)}$. Because we shall consider
 Yang-Mills theory later, we will keep $U/U^{-1}$ in $U(1)$ gauge transformation.

The Abelian gauge theory has the following specific properties which are different from 
the non-Abelian gauge theory: 

\noindent  (1) As long as the gauge anomalies are cancelled, the $U(1)$ charges for the particles 
can take arbitrary values, which give us the charge quantization problem in the SM.

\noindent (2) For two gauge theories $U(1)_I$ and $U(1)_{II}$, we can have the kinetic mixing term
$\kappa F^{I\mu\nu} F^{II}_{\mu\nu}$ since the field strength $F_{\mu \nu}$ is gauge invariant.

\noindent (3) The gauge anomalies can be cancelled by the Green-Schwarz mechanism 
or its generalization~\cite{Green:1984sg}.

\noindent (4) The gauge symmetry can be preserved while its gauge field becomes massive 
via Stueckelberg mechanism~\cite{Stueckelberg:1938zz}.

It seems to us that the third property might not be realized in the Yang-Mills theory since its field strength
is not gauge invariant. The last point is very interesting, but it is still a challenge question 
in the non-Abelian gauge theory. Therefore, in this paper, we shall try to realize the first two properties 
in the generalized non-Abelian gauge theory.

Next, we shall consider the generalized $U(1)$ gauge theory with multiple gauge Sectors.
For simplicity, we consider two gauge Sectors $I$ and $II$. The corresponding
$U(1)$ transformations for gauge fields,
scalars and fermions as well as the covariant derivative are 
\begin{eqnarray}
&& \Phi^{\alpha}_i \rightarrow U^{q_{\Phi^{\alpha}_i}} \Phi^{\alpha}_i~,~~
\Psi^{\alpha}_i \rightarrow U^{q_{\Psi^{\alpha}_i}} \Psi^{\alpha}_i~,~~\nonumber \\
&& A^{\alpha}_{\mu} \rightarrow U^{q_{A^{\alpha}}} A^{\alpha}_{\mu} (U^{q_{A^{\alpha}}})^{-1} 
+ \frac{i}{g_{\alpha}} U^{q_{A^{\alpha}}}\partial_{\mu} (U^{q_{A^{\alpha}}})^{-1}~,~\nonumber \\
&& D^{\alpha}_{\mu} = \partial_{\mu} -ig_{\alpha} \frac{q_{X^{\alpha}}}{q_{A^{\alpha}}} A^{\alpha}_{\mu}
 ~,~\,
\end{eqnarray}
where $\alpha = I,~II$.
As long as gauge anomalies are cancelled, we can introduce the mixed Sector where
the particles $X^M$ are couple to both $A^{I}$ and $A^{II}$ gauge fields. We need to introduce
a mixing angle $\theta_{X^M}$ for each particles, and the $U(1)$ transformation and covariant 
derivative are 
\begin{eqnarray}
&& \Phi^{M}_i \rightarrow U^{q_{\Phi^{M}_i}} \Phi^{M}_i~,~~
\Psi^{M}_i \rightarrow U^{q_{\Psi^{M}_i}} \Psi^{M}_i~,~~\nonumber \\
&& D^{M}_{\mu} = \partial_{\mu} -ig_{I} \frac{q_{X^{M}}}{q_{A^{I}}} c^2_{\theta_{X^M}} A^{I}_{\mu} 
 -ig_{II} \frac{q_{X^{M}}}{q_{A^{II}}} s^2_{\theta_{X^M}} A^{II}_{\mu} ~,~\, 
\end{eqnarray}
where $c^2_{\theta_{X^M}} \equiv \cos^2\theta_{X^M} $, and $s^2_{\theta_{X^M}} \equiv \sin^2\theta_{X^M} $.
For simplicity, we do not consider the mixed Sector here.

Redefining $\frac{q_{X^{\alpha}}}{q_{A^{\alpha}}}$ as $q_{X^{\alpha}}$ and using Abelian property, 
we obtain the following $U(1)$ transformation and covariant derivative 
\begin{eqnarray}
&& \Phi^{\alpha}_i \rightarrow U^{q_{\Phi^{\alpha}_i}} \Phi^{\alpha}_i~,~~
\Psi^{\alpha}_i \rightarrow U^{q_{\Psi^{\alpha}_i}} \Psi^{\alpha}_i~,~~\nonumber \\
&& A^{\alpha}_{\mu} \rightarrow U A^{\alpha}_{\mu} U^{-1} 
+ \frac{i}{g_{\alpha}} U\partial_{\mu} U^{-1}~,~\nonumber \\
&& D^{\alpha}_{\mu} = \partial_{\mu} -ig_{\alpha} q_{X^{\alpha}} A^{\alpha}_{\mu}~.~\,
\label{GABT-A}
\end{eqnarray}
Let us make the following rotation for two gauge fields
\begin{eqnarray}
\left(
\begin{array}{c}
A_\mu \\
{\widehat A}_\mu
\end{array} \right)=
\left(
\begin{array}{cc}
\cos\theta & \sin\theta \\
-\sin\theta & \cos\theta
\end{array}
\right)
\left(
\begin{array}{c}
{A}_\mu^{I} \\
{A}_\mu^{II}
\end{array} \right)
~,~\,
\label{RT-1}
\end{eqnarray}
where
\begin{eqnarray}
\sin\theta \equiv {{g_I}\over\displaystyle {\sqrt {g_I^2 +(g_{II})^2}}}~.~\,
\label{RT-2}
\end{eqnarray}
And then we obtain
\begin{eqnarray}
&& A_{\mu} \rightarrow U A_{\mu} U^{-1} 
+ i\frac{{\sqrt {g_I^2 +(g_{II})^2}}}{g_I g_{II}} U\partial_{\mu} U^{-1}~,~\nonumber \\
&& {\widehat A}_\mu \rightarrow U{\widehat A}_\mu  U^{-1}  ~.~\,
\label{RT-3}
\end{eqnarray}
Therefore, we can introduce the mass term for ${\widehat A}_\mu$ as follows
\begin{eqnarray}
V &=& \frac{1}{2} M_{\widehat A} ({\widehat A}_\mu)^2~.~\,
\label{RT-4}
\end{eqnarray}
For simplicity, we shall neglect it here. But, it can have rich physics as well. 
Of course, such rotation is not valid if two gauge fields obtain different masses
via Higgs mechanism.

In principle, if there is no interaction between two Sectors, we can consider
them as two independent $U(1)$ gauge symmetries. The non-trivial interactions between two Sectors break
two $U(1)$ gauge symmetries down to one $U(1)$ gauge symmetry. Thus, we obtain 
multiple gauge fields for one gauge transformation. The interesting property is
gauge symetry breaking. Let us present an simple example. We assume that
in Sector II, we have a Higgs field $\Phi^{II}$ with $U(1)$ charge 3, 
which acquire a Vacuum Expectation Value (VEV)
and break gauge symmetry. And then the gauge field $A^{II}_{\mu}$ becomes massive.
 In the Sector I, we introduce a Higgs field $\Phi_1^I$ with
$U(1)$ charge $-3$, and a Higgs field $\Phi_2^{I}$ with $U(1)$ charge $-1$.
The Higgs potential is given by
\begin{eqnarray}
V &=& m^2_{I1} |\Phi_1^{I}|^2 + \frac{\lambda_1}{2} |\Phi_1^{I}|^4 
+ m^2_{I2} |\Phi_2^{I}|^2 \nonumber \\ && 
+ \frac{\lambda_2}{2} |\Phi_2^{I}|^4 
 + m^2_{II} |\Phi^{II}|^2 + \frac{\lambda_3}{2} |\Phi^{II}|^4 + 
\nonumber \\ && 
+\left[ m_X^2 \Phi^{II}\Phi_1^I + \lambda_4 \Phi^{II} (\Phi_2^{I})^3 + {\rm H.C.}\right]~.~\,
\end{eqnarray}
Thus, if $m^2_{II} < 0$, $\Phi^{II}$ acquires a VEV. And then
 the Higgs field $\Phi_1^{I}$ will obtain a VEV via the tadpole term, and
the Higgs field $\Phi_2^{I}$ might get a VEV via the trilinear term. In particular, 
such trilinear term can give us strong first order phase
transition, which will have important applications in the SM for electroweak baryogenesis and
gravitation wave.
In short, after $\Phi^{II}$ gets the VEV, the $U(1)$ gauge symmetry is broken in
both Sectors. Moreover, we assume that in Sector I
 there are two fermions $\Psi^{I}_1 $ and $\Psi^{I}_2 $ with $U(1)$ charges $-1$ and $-2$.
And then we can have the Yukawa coupling term $y \Phi^{II} \Psi^{I}_1 \Psi^{I}_2$. After 
$\Phi^{II}$ obtains the VEV, we might still break the gauge symmetry in Sector I and 
give mass to gauge boson $A^{I}_{\mu}$ at one-loop level.

{\bf The Non-Abelian Gauge Theory and Its generalizations~--}~First, 
 we consider the simple generalization of the Yang-Mill theory with group $G$
and generator $T^a$, which have the scalars and fermions in the fundamental representation. It is simple to 
generalize the scalars and fermions in the generic representations.  Simialar to the above studies,
we consider two gauge Sectors I and II for simplicity, and
the correspoding $G$ transformation and covariant derivative are
\begin{eqnarray}
&& \Phi^{\alpha}_i \rightarrow U \Phi^{\alpha}_i~,~~
\Psi^{\alpha}_i \rightarrow U \Psi^{\alpha}_i~,~~\nonumber \\
 A^{\alpha}_{\mu} & \rightarrow & U A^{\alpha}_{\mu} U^{-1} 
+ \frac{i}{g_{\alpha}} U\partial_{\mu} U^{-1}~,~
F_{\mu\nu}^{\alpha} \rightarrow U F_{\mu\nu}^{\alpha} U^{-1} 
\nonumber \\
&& D^{\alpha}_{\mu} = \partial_{\mu} -ig_{\alpha}  A^{\alpha}_{\mu}~,~\,
\label{SGYM-A}
\end{eqnarray}
where $A^{\alpha}_{\mu} = A^{\alpha a}_{\mu} T^a$, and $U=e^{i\Lambda(x)}$ with
$\Lambda(x) = \Lambda^a(x) T^a $. 
Also, the gauge anomalies should be cancelled in each Sector. 
Similar to Eqs.~(\ref{RT-1}), (\ref{RT-2}), 
and (\ref{RT-3}), we can make the same tranformation for two kinds of gauge fields.
Also, we can introduce the following mass term for ${\widehat A}_\mu$ 
\begin{eqnarray}
V &=&  M_{\widehat A} {\rm Tr({\widehat A}_\mu)^2}~.~\,
\end{eqnarray}
For simplicity, we do not consider it here, although it can have rich physics as well. 
Moreover, we obtain that the kinetic mixing
between two kinds of gauge fields $-\frac{1}{\kappa} {\rm Tr}(F^{I\mu\nu} F^{II}_{\mu \nu})$
are gauge invariant.
For particles in the mixed Sector, the transformation and covariant derivative are 
\begin{eqnarray}
&& \Phi^{M}_i \rightarrow  \Phi^{M}_i~,~~
\Psi^{M}_i \rightarrow \Psi^{M}_i~,~~\nonumber \\
 D^{M}_{\mu} &=& \partial_{\mu} -ig_{I}  c^2_{\theta_{X^M}} A^{I}_{\mu} 
 -ig_{II}  s^2_{\theta_{X^M}} A^{II}_{\mu} ~.~\, 
\end{eqnarray}
Thus, these particles can have arbitrarily couplings to the
$A^I_{\mu}$ or $A^{II}_{\mu}$ gauge fields. In short, we indeed realize the first two specific properties
for the Abelian gauge theory. 

Next, we consider the generic generalization of the Yang-Mills theory. Similar to the above
discussions, we consider two Sectors as well as
the scalars and fermions in the fundamental representation.
The $G$ transformation and covariant derivative are
\begin{eqnarray}
&& \Phi^{\alpha}_i \rightarrow U^{q_{\alpha}} \Phi^{\alpha}_i~,~~
\Psi^{\alpha}_i \rightarrow U^{q_{\alpha}} \Psi^{\alpha}_i~,~~\nonumber \\
&& A^{\alpha}_{\mu} \rightarrow U^{q_{\alpha}} A^{\alpha}_{\mu} (U^{q_{\alpha}})^{-1} 
+ \frac{i}{g_{\alpha}} U^{q_{{\alpha}}}\partial_{\mu} (U^{q_{{\alpha}}})^{-1}~,~\nonumber \\
&& D^{\alpha}_{\mu} = \partial_{\mu} -ig_{\alpha} A^{\alpha}_{\mu}~,~~U^{q_{\alpha}}=e^{iq_{\alpha}\Lambda(x)}
 ~,~\,
\end{eqnarray}
where $q_{I} \not= q_{II}$. Without loss of generality, we can choose $q_{I}=1$, but we shall still
keep the generic formalism here.
In the mixed Sector, we can only have the fields in the tensor representations,
but not in the fundamental and anti-fundamental representations.
Also, the gauge anomalies should be cancelled in each Sector. 
Unlike the generalized Abelian 
gauge theory and the simple generlization of Yang-Mills theory, 
we do not have the kinetic mixing
between two kinds of gauge fields $-\frac{1}{\kappa} {\rm Tr}(F^{I\mu\nu} F^{II}_{\mu \nu})$,
and we cannot introduce
the mass term for any linear combination of two gauge fields except the Higgs mechanism 
with Higgs fields in the mixed sector.

For the interactions among the scalars and fermions in Sectors I and II, we need them to break the
$G\times G$ global symmetry down to $G$. Otherwise, we might equivalently have two gauge groups $G\times G$.
This is a challenge question in this kind of theory, and definitely deserve further detailed studies.
In order to have the interactions between two Sectors, we can introduce the bifundamental link fields
in the mixed Sector, or simply we can consider the product of two fields which are in the fundamental
and anti-fundametal representations respectively in two gauge Sectors. However, such kind of interactions 
will not break the additional global symmetry. Furthermore, 
suppose there is no interaction between two Sectors, we can choose different gauge transformations
for two Sectors. For an simple example, we can choose 
\begin{eqnarray}
U^I ~=~U^{q_{{II}}}~,~~~ U^{II} ~=~U^{q_{{I}}}~,~\, 
\end{eqnarray}
and thus we obtain the above simple generalization of the non-Abelian gauge theory given in Eq.(\ref {SGYM-A}).

{\bf Applications of the Generalized Gauge Theories in Particle Physics~--}~With the above 
generalized gauge theories, we shall consider their applications in the particle physics:

\noindent  (I) The Strong CP Problem and TeV-Scale Invisible Axion Models.
In the SM, we can introduce a CP-violating topological term for the $SU(3)_C$ gluon field strength 
\begin{eqnarray}
{\cal L} ~=~ {{\theta}\over {16\pi^2}} {\rm Tr} F_{\mu\nu} {\widetilde F}^{\mu \nu}~,~
~ {\widetilde F}^{\mu \nu} ~=~ \frac{1}{2} \epsilon^{\mu \nu \alpha \beta}F_{\alpha \beta} ~.~~ \, 
\end{eqnarray}
From the upper limits on the electric dipole moments of the neutron and $^{199}$Hg,
we obtain  $|\theta| <10^{-10}$. This fine-tuning problem is called strong CP problem,
and its natural solution is 
the PQ mechanism~\cite{Peccei:1977hh, Weinberg:1977ma, Wilczek:1977pj}. 
The main idea of PQ mechanism
is that we promote the $\theta$ parameter to a dynamic CP-odd Goldstone field axion from a global PQ
symmetry breaking. After QCD phase transition, the axion potential is generated via instanton effect,
and then we obtain $\theta =0$ by minimizing the potential and give mass to axion. 
Thus, the strong CP problem is solved.
Interestingly, axion can also be a cold dark matter candidate. Right now, there are two viable
invisible axion models: 
the Kim-Shifman-Vainstein-Zakharov (KSVZ) axion model~\cite{Kim:1979if, Shifman:1979if}, and 
the Dine-Fischler-Srednicki-Zhitnitsky (DFSZ) axion model~\cite{Dine:1981rt, Zhitnitsky:1980tq}. 
To be consistent with various 
experimental constraints and have the correct dark matter relic density, we need the PQ symmetry
breaking scale around $10^{11}$~GeV. An important question is why there exists an intermediate
scale for PQ symmetry breaking. However, we cannot construct the realistic TeV-scale axion models
due to the anomalous $U(1)_{\rm PQ}$ symmetry and experimental constraints. 
Interestingly, with the generalized non-Abelian 
gauge theory, we can indeed solve this problem.

We consider the simple generalization of the non-Abelian gauge theory, given in Eq.~(\ref{SGYM-A}).
We introduce two gauge fields $A_{\mu}^I$ and $A_{\mu}^{II}$ for $SU(3)_C$ gauge symmetry,
which are respectively the SM gluon fields and new gauge fields. In other words, the SM particles
belong to Sector I, and then $g_I=g_3$ in the SM model. In Sector I,
we also introduce a colored scalar $\widetilde{XD}_1$. 
In Sector II, we consider the KSVZ axion model,
and introduce a pair of vector-like  particles $(XT, ~XT^c)$,
the colored scalars $\Phi$ and $\widetilde{XD}_2$, as well as a SM singlet fermion $\chi$.
The $SU(3)_C\times SU(2)_L \times U(1)_Y$ quantum numbers for $\widetilde{XD}_i$,
$XT$, $XT^c$, $\Phi$, and $\chi$ are $(3, 1, -1/3)$, $(10, 1, 0)$, 
$(\overline{10}, 1, 0)$, $(6, 1, -2/3)$, and $(1, 1, 0)$, respectively.
Thus, the one-loop $SU(3)$ beta function in Sector II is zero. 
 Also, we introduce a SM singlet Higgs field $S$. To solve the strong CP problem, 
we introduce the $U(1)_{\rm PQ}$ symmetry under which the SM particles and $\widetilde{XD}_1$ are neutral. 
The $U(1)_{\rm PQ}$ charges
for $XT$, $XT^c$, $\Phi$, $\widetilde{XD}_2$, $\chi$  and $S$ are $1$, $1$, $1$, $1$,
$-1$, and $-2$, respectively.
The relevant Lagrangian is
\begin{eqnarray}
- {\cal L} &=& \kappa {\rm Tr}(F^{I\mu\nu} F^{II}_{\mu \nu}) + V(S, ~\Phi,~\widetilde{XD}_i)
+ \left( \lambda_T S XT^c XT 
\right. \nonumber \\ && \left.
+ y^u_i XT \Phi^* U_i^c + A_{XD} \Phi^* \widetilde{XD}_1 \widetilde{XD}_2 
+ y^d_i \widetilde{XD}_2 D_i^c \chi 
\right. \nonumber \\ && \left. 
+ y_{ij} \widetilde{XD}^*_1 U_i^c D_j^c +y^q_{ij} \widetilde{XD}^*_1 Q_i L_j 
+ \lambda_{\chi} S^* \chi\chi +{\rm H. C.}\right) ~,~~ \nonumber\, 
\end{eqnarray}
where $Q_i$, $L_i$, $U_i^c$ and $D_i^c$ are respectively the left-handed quark doublet, lepton doublet,
 right-handed up-type and down-type quarks, and
the scalar potential is 
\begin{eqnarray}
 V &=& m_S^2 |S|^2 +\frac{\lambda_S}{2} |S|^4  + m_X^2 |X|^2 
+\frac{\lambda_X}{2} |X|^4~,~~
\end{eqnarray}
where $X= \Phi, ~\widetilde{XD}_1$, and $\widetilde{XD}_2$.
There is one and only one $SU(3)_C$ gauge transformation due to the 
$y^u_i$, $A_{XD}$, and $y_i^d$ terms.
To avoid the proton decay, we forbid the $y^q_{ij}$ term by lepton number conervation.
After $S$ acqiures a VEV due to negative $m_S^2$, the $U(1)_{\rm PQ}$ symmetry is broken. The complex phase of $S$ is
the Nambu-Goldstone boson, {\it i.e.}, axion. The axion obtains a mass after QCD phase transition
due to instanton effect. In the original KSVZ axion model, the VEV of $S$ is the one and only one parameter, 
{\it i.e.}, the axion decay constant $f_a$.
In our model, we obtain
\begin{eqnarray}
{\cal L} ~=~ {{15}\over {16\pi^2}} \frac{a}{f_a} {\rm Tr} F_{\mu\nu} {\widetilde F}^{\mu \nu}~,~ \, 
\end{eqnarray}
where $a = {\rm Im } S/{\sqrt 2}$, and the axion decay constant is
\begin{eqnarray}
f_a = \frac{15g_{II}^2 \kappa^2}{g_I^2 \langle S \rangle}=\frac{15 g_{II}^2 \kappa^2}{g_3^2 \langle S \rangle} ~.~\,
\end{eqnarray}
Assuming $g_{II} \kappa \simeq 10^{-4} g_{3}/\sqrt{15}$ as well as $\langle S \rangle \simeq 1$~TeV,
we obtain $f_{a} \simeq 10^{11}$~GeV. For example, we can choose
$g_{II} \simeq g_{3}$ and $ \kappa \simeq 10^{-4}/\sqrt{15}$, or $g_{II} \simeq 0.1 \times g_{3}$ and 
$ \kappa \simeq 10^{-3}/\sqrt{15}$.
Therefore, we indeed construct the TeV-scale invisible axion models.
Furthermore, we can introduce a $Z_2$ symmetry under which $XT^c$, $XT$,  $\Phi$, 
$\widetilde{XD}_2$, and $\chi$ are odd while
all the other particles are even. And then $\chi$ can be a dark matter candidate. 

We would like to emphasize that to escape the experimental constraints,
the vector-like quarks $XT$ and $XT^c$ must be $SU(2)_L\times U(1)_Y$ singlets.
The simplest axion model is to introduce the vector-like quarks $XD0^c$ and $XD0$ with
SM quantum numbers $({\bar 3}, 1, 0)$ and $(3, 1, 0)$ inspired 
from string model building~\cite{Dienes:1996du, Blumenhagen:2005mu, Barger:2007qb} in Sector II.
Because these particles are stable, to avoid the stringent experimental constraints, 
we require that the reheating temperature be smaller than their mass, {\it i.e.},
we need low reheating temperature~\cite{Barger:2007qb}.
The detailed studies will be given elsewhere~\cite{TL-preparation}.

\noindent  (II) Higgs Mechanism. We consider the simple generalization of the $SU(2)_L\times U(1)$
gauge symmetry for two gauge sectors with gauge transformations given 
in Eqs.~(\ref{GABT-A}) and (\ref{SGYM-A}).
 In Sector I we have the SM, {\it i.e.}, $W^{Ia}_{\mu}$ and $B^I_{\mu}$ are
the SM gauge bosons, and the SM fermions and Higgs field $H^I$ belong to this Sector as well.
In Sector II, we introduce Higgs field $H^{II}$. The $SU(3)_C\times SU(2)_L\times U(1)_Y$ quantum numbers 
for $H^I$ and $H^{II}$ are all $(1, 2, -\frac{1}{2})$. The Higgs potential is 
\begin{eqnarray}
V &=&  m^2_{I} |H^{I}|^2 + \frac{\lambda_1}{2} |H^{I}|^4 
+ m^2_{II} |H^{II}|^2 + \frac{\lambda_2}{2} |H^{II}|^4 
\nonumber \\ && 
+ \lambda_3 |H^{I}|^2 |H^{II}|^2
+\left( m_X^2  {\widetilde H}^I
 H^{II} + \lambda_4 {\widetilde H}^I  H^{II} |H^I|^2 
\right. \nonumber \\ && \left.
+ \lambda_5 {\widetilde H}^I  H^{II} |H^{II}|^2 + \lambda_6 ({\widetilde H}^I  H^{II})^2
 + {\rm H.C.}\right)~,~\,
\end{eqnarray}
where ${\widetilde H}^I = i \sigma_2 H^{I*}$. With $m^2_{II}<0$, we break
the gauge symmetry in Sector II and $H^{II}$ acquires a VEV. And then
we obain a tadpole term and a trilinear term for $H^I$. Thus, the SM electroweak gauge symmetry
will be broken and the strong first-order electroweak phase transition can realized 
in Sector I, which is important for electroweak baryogenesis and gravitational wave.

\noindent  (III) The deconstructions of extra dimensions~\cite{ArkaniHamed:2001ca, Hill:2000mu} can be realized by the
generalized Abelian and non-Abelian gauge theories with $N$ gauge sectors.

\noindent  (IV)  The Mirror models~\cite{Lee:1956qn, Kobzarev:1966qya, Pavsic:1974rq, Kolb:1985bf, Foot:1991bp},
top color models~\cite{Topcolor}, top flavour models~\cite{Topflavor}, 
and top hypercharge models~\cite{Chiang:2007sf} can be considered as the SM gauge symmetries
with two gauge Sectors, and our proposal might be applied to the heterotic $E_8\times E_8$
string theory~\cite{Gross:1984dd, Gross:1985fr, Horava:1995qa, Horava:1996ma}.

{\bf  Discussions and Conclusion~--}~Assuming that there are multiple gauge sectors 
for one gauge transformation, 
we studied the generalized Abelian and non-Abelian gauge theories. We first 
pointed out that the $U(1)$ gauge theory has a few unique properties, which 
provide the motivations for the generalized Yang-Mills theory. Also, we considered 
the generalized Abelian gauge theory, and studied the Higgs mechanism. In addition, 
we proposed the simple and generic generalizations of Yang-Mills theory. 
In the simple generalization, we realize two specific properties in the Abelian 
gauge theory.  For the applications in particle physics, we proposed
the TeV-scale invisible axion model which can still solve the strong CP problem, 
and has the effective decay constant around the intermediate scale. And we studied 
the SM electroweak gauge symmetry breaking induced from the symmetry breaking in the 
other gauge Sector. In particular, we can easily obtained the strong first order 
electroweak phase transition in the SM, which is important for electroweak baryogenesis 
and gravitational wave. We comment the relevant model building as well.

{\bf Acknowledgments~--}~We would like to thank Stephen M. Barr and Pei-Hong Gu 
for helpful discussions.
This research was supported by
the Projects 11475238 and 11647601 supported by the 
National Natural Science Foundation of China, and by 
the Key Research Program of Frontier Science, CAS.


\end{document}